\documentclass[twocolumn,showpacs,preprintnumbers,amsmath,amssymb,superscriptaddress]{revtex4}

\divide\topmargin by 3 

\usepackage{graphicx}
\usepackage{dcolumn}
\usepackage{bm}

\usepackage{amsmath}
\newcommand{\eV}{\rm eV}
\begin{document}
\title{Anomalous Bias Dependence of Spin Torque in Magnetic Tunnel Junctions}
\author{Ioannis Theodonis} \affiliation{Department of Physics, California
State University, Northridge, CA 91330-8268} \affiliation
{Department of Physics, National Technical University, GR-15773,
Zografou, Athens, Greece}
\author{Nicholas Kioussis}
\email[E-mail me at: ]{nick.kioussis@csun.edu.}
 \affiliation{Department of Physics,
California State University, Northridge, CA 91330-8268}
\author{Alan Kalitsov}
\affiliation{Department of Physics, California State University,
Northridge, CA 91330-8268}
\author{Mairbek Chshiev}
\affiliation{MINT Center, University of Alabama, P.~O.~Box 870209,
Tuscaloosa, AL, USA}
\author{W. H. Butler}
\affiliation{MINT Center, University of Alabama, P.~O.~Box 870209,
Tuscaloosa, AL, USA}
\date{\today}
\begin{abstract}
{We predict an anomalous bias dependence of the spin transfer torque
parallel to interface, $T_{\|}$, in magnetic tunnel junctions (MTJ),
which can be selectively tuned by the exchange splitting. It may
exhibit a sign reversal {\it without} a corresponding sign reversal
of the bias or even a quadratic bias dependence. We demonstrate that
the underlying mechanism is the interplay of spin currents for the
ferromagnetic (antiferromagnetic) configurations, which vary
linearly (quadratically) with bias, respectively, due to the
symmetric (asymmetric) nature of the barrier. The spin transfer
torque perpendicular to interface exhibits a quadratic bias
dependence.}
\end{abstract}
 \maketitle

Theoretical calculations predict that when a spin-polarized current
passes through a magnetic multilayer structure, whether spin
valve\cite{Slonczewski} or magnetic tunnel junction,
(MTJ)\cite{Slonczewski1,Slonczewski2}, it can transfer spin angular
momentum from one ferromagnetic electrode to another, and hence
exert a torque on the magnetic moments of the electrodes. At
sufficiently high current densities, this spin transfer can
stimulate spin-wave excitations\cite{Tsoi,Krivorotov} and even
reverse the magnetization of an individual domain\cite{Myers}.
Current-induced magnetic switching (CIMS) has now been confirmed in
numerous experiments both in spin valves\cite{Myers,katine} and more
recently, in MTJs\cite{Fuchs,fuchs2}. Thus, CIMS provides a powerful
new tool for the study of spin transport in magnetic nanostrutures.
In addition, it offers the intriguing possibility of manipulating
high-density nonvolatile magnetic-device elements, such as
magnetoresistive random access memory (MRAM), without applying
cumbersome magnetic fields\cite{parkin}.

While the fundamental physics underlying the spin transfer torque
(STT) in spin valves has been extensively studied
theoretically\cite{Slonczewski,bauer,mathon,stiles}, its role in
MTJs remains an unexplored area thus far, except for the pioneering
work of Slonczewski~\cite{Slonczewski1,Slonczewski2}, who employed
the free-electron model in the low bias regime. One of the most
pressing needs is a comprehensive understanding of the bias
dependence of the STT in MTJs, which will be important for the
development of MRAM that uses CIMS for writing the magnetic memory
cell.

In this Letter, we present for the first time a comprehensive study
of the effect of bias on the spin torques, parallel ($T_{\|}$) and
perpendicular ($T_{\bot}$) to the interface, in MTJs, using
tight-binding (TB) calculations and the non-equilibrium Keldysh
formalism. We predict an anomalous bias dependence of the spin
torque, contrary to the general consensus. We demonstrate first that
depending on the exchange splitting, $T_{\|}$ may exhibit an unusual
 non-monotonic bias dependence: it may change sign
{\it without a sign reversal in bias or current}, and it may even
have a quadratic bias dependence. Second, by generalizing the
equivalent circuit in Ref. \cite{Slonczewski2} using
angular-dependent resistances, we show that $T_{\|}$ satisfies an
expression involving the {\it difference} in spin currents between
the ferromagnetic (FM) and antiferromagnetic (AF) configurations.
This expression is general and {\it independent} of the details of
the electronic structure. Our numerical results both for the TB
model and the free-electron model (not presented here) confirm the
validity of this relation for any parameter set and bias. Third, the
spin current for the FM (AF) alignment is shown to have a linear
(quadratic) bias dependence, whose origin lies in the symmetric
(asymmetric) nature of the barrier. The interplay of the spin
currents for the FM and AF configurations is the key underlying
mechanism that can lead to a rich behavior of the STT on bias.
Finally, we find that the bias dependence of $T_{\bot}$  is
quadratic.

\begin{figure}
\begin{center}
\includegraphics[width=7.5cm]{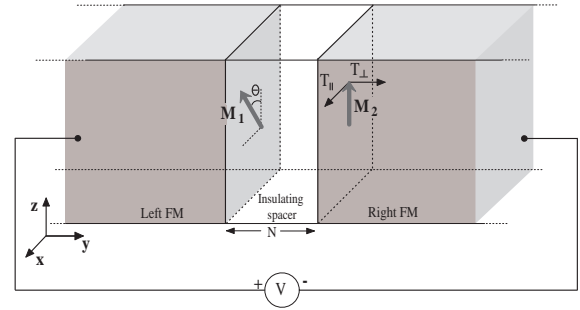}
\caption{\label{fig1}Schematic structure of the MTJ, consisting of
left and right semi-infinite FM leads separated by a thin
nonmagnetic insulating system containing N atomic layers. The
magnetization ${\bf M}_2$ of the right FM lead is along the z-axis,
whereas the magnetization ${\bf M}_1$ of the left lead is rotated by
angle $\theta$ around the $y$-axis with respect to  ${\bf M}_2$.}
\end{center}
\end{figure}

The MTJ under consideration depicted in Fig.1 consists of a left and
right semi-infinite noncollinear FM leads, separated by a
nonmagnetic insulating (I) spacer containing $N$ atomic layers. The
right FM lead is magnetized along the z axis ({$\bf M_2$}) of the
coordinate system, shown in the inset of Fig. 1. The magnetization
of the left FM lead (${\bf M_1}$) lies in the $x-z$ interfacial
plane, i.e. it is rotated by angle $\theta$ around the axis $y$
(normal to the FM/I interfaces) with respect to ${\bf M_2}$. The
chemical potentials of the right and left leads are shifted by the
external bias, $\eV = \mu_L - \mu_R$, where the charge current is
positive when it flows along the $y$ axis from left to right.
The Hamiltonian for each region is described by a single orbital
simple-cubic TB model with a nearest-neighbor (NN) spin-independent
hopping term, $t_{\alpha}$, and a spin-dependent on-site energy
term, $\varepsilon_{\alpha}^{\sigma}$, where $\alpha$= L, R, and I,
refer to the left, right, and insulating regions, respectively. In
the present calculations the left and right FM leads are identical
with an exchange-splitting, $\Delta_{\rm L(R)}=\varepsilon_{\rm
L(R)}^\uparrow-\varepsilon_{\rm {L(R)}}^\downarrow$. We use
$\Delta_{\rm I}=0$, $\varepsilon_{\rm L(R)}^\uparrow-E_F=1.2\eV
$, $\varepsilon_{\rm I}-E_F=5.4\eV$, $N=5$, and $t_{\rm R(L)}=t_{\rm I}
=t_{\rm R(L),\rm I}=t=0.4~\eV$, where the Fermi
energy $E_F=0 eV$, and the $t_{\rm L,I}$ and $t_{\rm R,I}$ are the
NN hopping matrix elements at the two FM/I interfaces~\cite{itoh}.
Note, that $\Delta_{\rm L(R)}$, refers to the {\it local effective
s-d} exchange interaction,
with values between 0.2 eV to 2 eV~\cite{sanvito} depending on the material.
Under applied bias, $\varepsilon_{\rm R}^{\sigma} - \varepsilon_{\rm
L}^{\sigma}= \eV$, the potential inside the insulator,
$\varepsilon_{{\rm I},n}=\varepsilon_{\rm I} - {\eV}\frac{n-1}{N-1}$
varies linearly with layer number $n$.

The one-electron Schr\"odinger equation in spin space for each
uncoupled region $\alpha$ is\cite{kalitsov}
\begin{multline}
\sum_{\scriptstyle p_1 \atop \scriptstyle p,q,p_1 \in \alpha}
\displaystyle\biggl\{ \bigl[(E-\varepsilon_{{\bf
k}_{||}})\delta_{pp_1}-\bar{H}_{pp_{1}}\bigr]\widehat{I}-
\delta
H_{pp_{1}} \\
 \times \left (
\begin{array} {ccc} cos\theta & sin\theta \\
sin\theta & -cos\theta \end{array} \right ) \displaystyle\biggr\}
\left(
\begin{array} {ccc} g_{p_{1}q}^{\uparrow \uparrow} &
g_{p_{1}q}^{\uparrow \downarrow} \\ g_{p_{1}q}^{\downarrow \uparrow}
& g_{p_{1}q}^{\downarrow \downarrow} \\ \end{array} \right) =
\delta_{pq} \widehat{I},\label{hamiltonian}
\end{multline}
where $p$ and $q$ are atomic sites indices in region $\alpha$,
$\varepsilon_{{\bf k}_{||}}$ is the energy of the in-plane wave
vector, ${\bf k}_{||}$, of the Bloch state, $g_{pq}^{\sigma\sigma
'}$ is the spin-dependent retarded Green's function for each region,
 and $\widehat{I}$ is the $2\times2$ unit matrix. The quantities
$\bar{H}_{pq}=\frac{1}{2} [\left ( \varepsilon_{\alpha}^{\uparrow} +
\varepsilon_{\alpha}^{\downarrow} \right)\delta_{pq} +
t_{\alpha}\left(\delta_{p,q+1}+\delta_{p,q-1}\right)]$, and $\delta
H_{pq}=\frac{1}{2}\Delta_{\alpha}$, describe the spin-average and
the spin-split part of the Hamiltonian, respectively.

Having determined the $g_{pq}^{\sigma,\sigma '}$ for each uncoupled
subsystem from Eq.~(\ref{hamiltonian}), one can calculate the
retarded Green's function for the entire coupled system, by solving
a system of Dyson equations which couples the $g_{pq}^{\sigma,\sigma
'}$ through the hopping matrix elements $t_{L,I}$ and  $t_{I,R}$ at
the two interfaces. In order to calculate the non-equilibrium
Keldysh Greens function, $\widehat{G}^{< \ \sigma,\sigma
'}_{i,i+1}$ we have extended the approach of Caroli {\it et
al.}~\cite{Caroli} in spin space using  $2\times2$  Greens function matrices.

The charge current density is~\cite{mathon}
\begin{equation}
I =I^{\uparrow}+I^{\downarrow}= \frac{e t}{2\pi\hbar} \int
Tr_{\sigma} [\widehat{G}^{< \ \sigma,\sigma '}_{i,i+1} -
\widehat{G}^{< \ \sigma,\sigma '}_{i+1,i}]dE d{\bf k}_{||} ,
\label{current}
\end{equation}
and the spin current density between sites i and i+1 is
\begin{equation}
\mbox{\boldmath$I$}^{(s)}_{i,i+1} = \frac{t}{4\pi} \int Tr_\sigma
\left [ (\widehat{G}^{< \sigma,\sigma '}_{i,i+1} - \widehat{G}^{<
\sigma,\sigma '}_{i+1,i})\mbox{\boldmath$\sigma$} \right ] dE d{\bf
k}_{||} \label{spincurrent},
\end{equation}
where $\mbox{\boldmath$\sigma$}=(\sigma_x,\sigma_y,\sigma_z)$ is a
vector of the Pauli matrices. Both $I^{\uparrow}$ and
$I^{\downarrow}$ are conserved across the MTJ, while the spin
current is not conserved (${\bf
\nabla}\cdot\mbox{\boldmath$I$}^{(s)}\neq 0$), due to spin-dependent
scattering caused by the local exchange field inside the FM
leads~\cite{stiles}. Conservation of the total angular momentum
implies that the spin current lost at an atomic site is transferred
to its local magnetic moment, thereby exerting a local STT
~\cite{stiles,mathon} ${\bf T}_i$ on site $i$ in the right FM lead
per unit area defined by
\begin{equation}
{\bf T}_i \equiv -\nabla\cdot\mbox{\boldmath$I$}^{(s)}=\mbox{\boldmath$I$}^{(s)}_{i-1,i}-
\mbox{\boldmath$I$}^{(s)}_{i,i+1}, \label{eq4}
\end{equation}
where the second equality represents the discrete form of the
divergence of the spin current. The local spin torque, ${\bf T}_i$,
shown in Fig. 1, has components parallel $(T_{i,\|})$ and
perpendicular $(T_{i,\bot})$ to the interfacial plane.  The
z-component of ${\bf T}_i$ vanishes because $I^{(s)}_{(i,i+1),z} =
I^{(s)}_{(i-1,i),z} =\frac{\hbar}{2e}(I^{\uparrow}-I^{\downarrow})$.
Both $T_{i,\|}$ and $T_{i,\bot}$ oscillate with different phase and
decay with distance from the I/FM interface, as in the case of spin
valves~\cite{bauer,stiles,kalitsov}.
\begin{figure}
\begin{center}
\includegraphics[width=8.5cm]{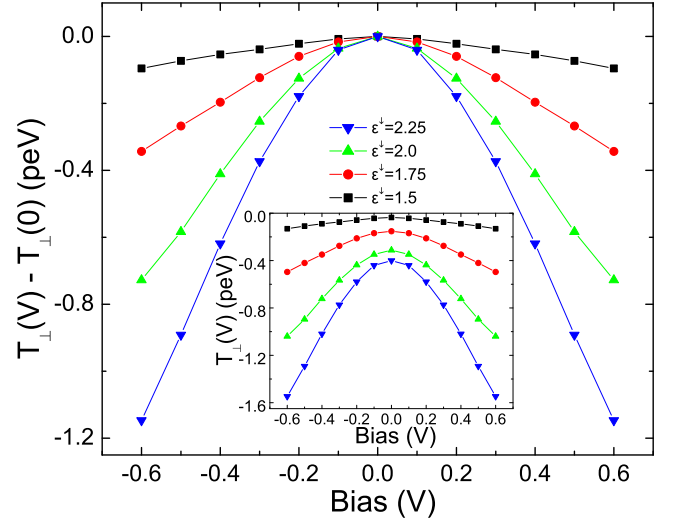}
\caption{\label{fig2}(Color Online) Bias dependence of the
current-induced
 perpendicular component of the {\it net}
spin torque per unit area, $T_{\bot}(V)-T_{\bot}(0)$, for $\theta =
\pi/2$, and various values of $\varepsilon^{\downarrow}$.
$T_{\bot}(0)$ is related to the exchange coupling energy between
left and right FM leads. Inset: bias dependence of $T_{\bot}(V)$. }
\end{center}
\end{figure}

The {\it net} STT transverse to the magnetization on the right FM
lead, a quantity which presumably experiment measures, is the sum of
local torques, ${\bf T}_i$,
\begin{equation}
{\bf T}=\sum_{i=0}^{\infty}(\mbox{\boldmath$I$}^{(s)}_{i-1,i}-
\mbox{\boldmath$I$}^{(s)}_{i,i+1})=
\mbox{\boldmath$I$}^{(s)}_{-1,0}-\mbox{\boldmath$I$}^{(s)}_{\infty,\infty} =
\mbox{\boldmath$I$}^{(s)}_{-1,0},
\end{equation}
where the subscripts -1 and 0 refer to the last site inside the
barrier and the first site in the right FM lead, respectively. In
the above equation, $\mbox{\boldmath$I$}^{(s)}_{\infty,\infty} = 0$ because the
components of $\mbox{\boldmath$I$}^{(s)}_{i,i+1}$ transverse to ${\bf M}_2$ decay to
zero as $i\rightarrow \infty$~\cite{bauer}. Thus, the {\it net} spin
torque exerted on the right FM lead is simply the spin current
calculated at the I/FM interface~\cite{stiles}.
\begin{figure}
\begin{center}
\includegraphics[width=8.5cm]{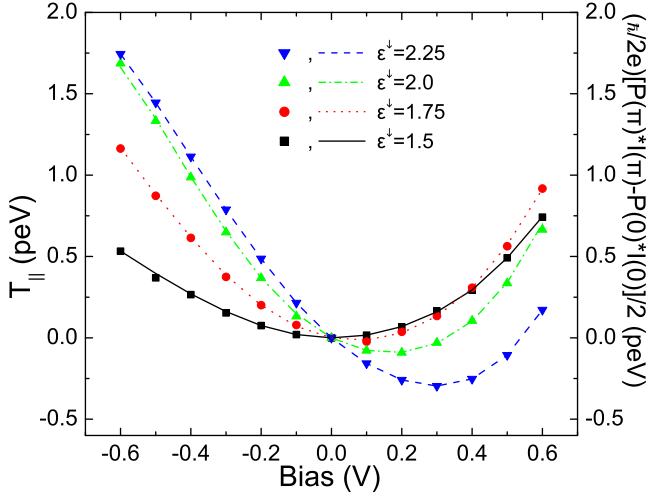}
\caption{\label{fig3}(Color Online) Bias dependence of the parallel
component of the {\it net} spin transfer torque per unit area,
$T_{||}$, for $\theta = \pi/2$, and various values of
$\varepsilon^{\downarrow}$. The curves (symbols) refer to the left-
(right-) hand ordinate, respectively.}
\end{center}
\end{figure}

In Fig.~\ref{fig2} we show the bias-dependence of the
current-induced perpendicular component,
$T_{\bot}(V)-T_{\bot}(V=0)$, of the {\it net} spin torque for
$\theta=\pi/2$, and various values of $\varepsilon^{\downarrow}$.
The inset displays the bias dependence of $T_{\bot}(V)$. We find
that $T_{\bot}(V)$ varies quadratically with bias, as originally
suggested, but not calculated, by Slonczewski~\cite{Slonczewski2}.
The equilibrium ($V=0$) value of $T_{\bot}(V)$, related to the
interlayer exchange coupling energy~\cite{Slonczewski}, decreases in
absolute value as the exchange splitting $\Delta$ is reduced.

In Fig.~\ref{fig3} we display the bias dependence of the parallel
component of the {\it net} spin torque, $T_{||}$,  (curves
associated with the left ordinate) for $\theta=\pi/2$ and for the
same values of $\varepsilon^{\downarrow}$ as in Fig.~\ref{fig2}. The
most striking and surprising feature of $T_{||}$ is its
non-monotonic bias dependence, which can vary from almost linear to
purely quadratic behavior, depending on the exchange splitting
$\Delta$. The quadratic bias dependence of $T_{||}$ for
$\varepsilon^{\uparrow}=1.5~\eV$ persists even for small bias.
Interestingly, for $\varepsilon^{\downarrow}=2~\eV$ and $2.25~\eV$,
$T_{||}$ reverses its sign {\it without a sign reversal in bias or
current.} This anomalous bias behavior may have important practical
implications, since it suggests that the CIMS in MTJ may not require
reversal of the current.  Note, that $T_{\bot}$ and $T_{||}$, are
comparable in size in MTJ, in contrast to metallic spin valves,
where $T_{\bot} \ll T_{||}$~\cite{bauer}.

\begin{figure}
\begin{center}
\includegraphics[width=8.0cm]{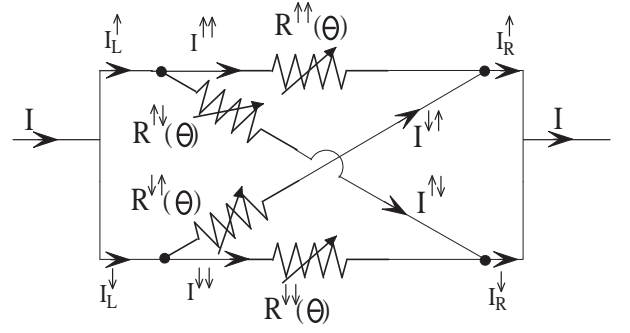}
\caption{\label{fig4} Equivalent circuit for spin-channel currents,
with angular-dependent resistances.}
\end{center}
\end{figure}

In order to understand the underlying mechanism responsible for the
bias-dependence of $T_{||}$, we have generalized the equivalent
circuit for MTJ~\cite{Slonczewski2}, using angular-dependent
resistances, $R^{\sigma,\sigma}(\theta)= R^{\sigma}(0)cos
^{-2}(\theta/2)$ and $R^{\sigma,\overline{\sigma}} (\theta)=
R^{\sigma}(\pi)sin ^{-2}(\theta/2)$, as displayed in Fig.
\ref{fig4}. The angular dependence of $R^{\sigma,\sigma'}(\theta)$
is equal to the inverse probability
$[P^{\sigma,\sigma'}(\theta)]^{-1}$ for an electron with spin state
$|\sigma>$ quantized along ${\bf M}_1$ to tunnel to a spin state
$|\sigma'>$ quantized along ${\bf M}_2$, where multiple reflections
within the barrier are neglected\cite{theodonis2}.
Substituting the currents $I^{\sigma}_{L(R)}$ in Fig. 4 into Eq. (5)
of Ref.~\cite{Slonczewski2}, we obtain
\begin{equation}
T_{||}(\theta)=\frac{I^{(s)}_z(\pi)- I^{(s)}_z(0)}{2} {\bf
M}_{2} \times ({\bf M}_{1}\times {\bf M}_{2}), \label{correlation}
\end{equation}
where
$I^{(s)}_z(\pi)=\frac{\hbar}{2e}(I^{\uparrow}(\pi)-I^{\downarrow}(\pi))$
and
$I^{(s)}_z(0)=\frac{\hbar}{2e}(I^{\uparrow}(0)-I^{\downarrow}(0))$
are the spin-current densities
for the AF and FM configurations, respectively. This important result
is quite general, {\it independent} of the details
of the electronic structure, and reduces the calculation of
$T_{||}(\theta)$ simply to the
 evaluation of the spin-current densities for the FM and
AF configurations~\cite{Slonczewski2}. The angular dependence of
both $T_{\bot}$ and $T_{||}$ is proportional to
$\sin\theta$~\cite{Slonczewski1}, in contrast to metallic spin
valves~\cite{Slonczewski,bauer}.

Defining the dynamic current polarization~\cite{tinkhman},
$P(\theta)=
[I^{\uparrow}(\theta)-I^{\downarrow}(\theta)]/I(\theta)$, where
$I(\theta)=I^{\uparrow}(\theta)+I^{\downarrow}(\theta)$ is the total
current, Eq.~(\ref{correlation}) reduces to
\begin{equation}
T_{||}(\theta)=\frac{\hbar}{2e}\frac{P(\pi)I(\pi)-P(0)I(0)}{2}
{\bf M}_{2} \times ({\bf M}_{1}\times {\bf M}_{2}).
\label{correlation2}
\end{equation}
In order to confirm Eq.~(\ref{correlation}) or
Eq.~(\ref{correlation2}) we display also in Fig.~\ref{fig3} the bias
dependence of $\frac{\hbar}{2e}[P(\pi)I(\pi)-P(0)I(0)]/2$ (symbols
associated with the right ordinate) for the same values of
$\varepsilon^{\downarrow}$, where the agreement is excellent.

\begin{figure}
\begin{center}
\includegraphics[width=8.5cm]{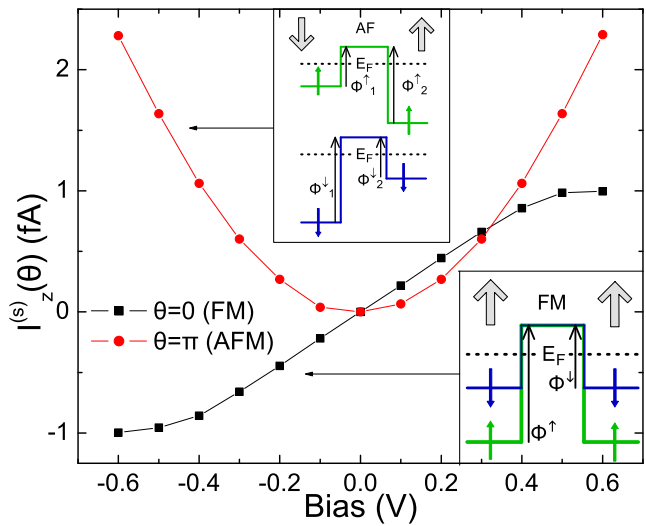}
\caption{\label{fig5}(Color Online) Bias dependence of the
spin-current density, $I^{(s)}_z (\theta)$, for the FM and AF
orientations, respectively. In the FM case the majority and minority
electrons tunnel through a symmetric barrier (lower inset) with
different barrier heights, $\Phi^\sigma$, from the bottom of the
band. In the AF case (upper inset), the two spin channels tunnel
through asymmetric barriers with the same average barrier height,
$\bar\Phi^\sigma$, but with a barrier asymmetry,
$\Delta\Phi^\sigma$, of opposite sign.}
\end{center}
\end{figure}
In order to elucidate the atomistic origin of the bias dependence of
$T_{||}$ in Eq.~(\ref{correlation}), we plot in Fig.~\ref{fig5} the
spin-current densities, $I^{(s)}_z(\pi)$ and $I^{(s)}_z(0)$,  versus
bias for the AF and FM orientations, respectively for
$\varepsilon^\downarrow =2.0~\eV $. One can clearly see that
$I^{(s)}_z(0)$ ($I^{(s)}_z(\pi)$) varies linearly (quadratic) with
bias for $V < 0.5~\eV$. For the FM and AF configurations the
tunneling can be considered as the superposition of two independent
spin channels. This different bias behavior can be understood on the
basis of the tunnel model~\cite{brinkman} for asymmetric barriers,
generalized so as to take into account both spin channels. The bias
dependence of $I^{\sigma}$ can be written as~\cite{brinkman},
$I^{\sigma} ({\rm V}) = f_1(\bar\Phi^{\sigma}){\rm V} -
f_2(\bar\Phi^{\sigma})\Delta\Phi^\sigma{\rm V}^2 + O({\rm V}^3), $
where $f_1$ and $f_2$ are functions of the average barrier height,
$\bar\Phi^\sigma = [\Phi_1^\sigma+\Phi_2^\sigma]/2$, and
$\Delta\Phi^\sigma = \Phi_1^\sigma - \Phi_2^\sigma$ is the barrier
asymmetry. Here, $\Phi_{1(2)}^\sigma$ is the spin-dependent barrier
height at the left (right) interface.

In the FM configuration, the majority and minority electrons tunnel
through a {\it symmetric} barrier (lower inset in Fig.~\ref{fig5})
but with different barrier heights, $\Phi^\sigma$, for each spin
channel. In this case, $\bar{\Phi}^{\uparrow} \neq
\bar{\Phi}^{\downarrow}$ and $\Delta\Phi^{\uparrow} =
\Delta\Phi^{\downarrow}=0$. Thus, both $I^\uparrow (0)$ and
$I^\downarrow(0)$ vary linearly with ${\rm V}$, and hence also
$I^{(s)}_z(0)$. On the other hand, in the AF configuration (upper
inset in Fig.~\ref{fig5}), both spin channels tunnel through
asymmetric barriers with the same average barrier height,
$\overline{\Phi}^{\uparrow} = \overline{\Phi}^{\downarrow}$, but
with barrier asymmetry of opposite sign, $\Delta\Phi^{\uparrow} = -
\Delta\Phi^{\downarrow}$. Hence, the linear bias dependence of
$I^\uparrow (\pi)- I^\downarrow(\pi)$ vanishes identically, and
$I^{(s)}_z(\pi)$
 exhibits a quadratic bias dependence.

Thus, the {\it interplay} between the linear and quadratic
bias-dependence of $I^{(s)}_z(0)$ and $I^{(s)}_z(\pi)$,
respectively, in Eq.~(\ref{correlation}) is responsible for the
non-monotonic bias-dependence of $T_{||}$ in Fig.~\ref{fig3}. This
competition can be selectively tuned by varying $\Delta$, giving
rise to a wide range of rich bias behavior. For example, the purely
quadratic bias behavior for $\varepsilon^{\downarrow}=1.5~\eV$
 arises from the fact that $I^\uparrow(0)= I^\downarrow(0)$.
Contrary to the free-electron model~\cite{Slonczewski1}, this
behavior is possible within the TB model due to the bell-like form
of the density of states at the interfaces~\cite{theodonis2}.

In summary, we predict an anomalous bias behavior of $T_{\|}$ in
MTJ, which varies with the exchange splitting in the FM leads.
$T_{\|}$ may exhibit a sign reversal {\it without} a corresponding
sign reversal of the bias, or even an unexpected quadratic bias
dependence. The underlying mechanism for the unusual bias dependence
is the interplay between the bias dependence of the spin currents
for the FM and AF configurations. The origin for the linear
(quadratic) bias dependence of the spin currents is the symmetric
(asymmetric) nature of the tunnel barrier for the FM (AF)
orientations. We should emphasize that the non-monotonic bias
behavior is not associated to the simple TB model; other systems
with more complex electronic structures can also show this behavior,
provided that the condition $I^{(s)}_z(0) < I^{(s)}_z(\pi)$ is
satisfied. On the other hand, $T_{\bot}$
exhibits a quadratic bias dependence.

 An experimental test of our
prediction can be achieved by relating the (observable) critical
voltage for switching measured as a function of external magnetic
field to the spin-torque. This can be done for a magnetic element
that has been characterized as to damping factor, anisotropy, and
magnetic moment. Future work will be aimed to include  the results
of these calculations as an input into the Landau-Lifshitz-Gilbert
equation, to calculate the critical current for the CIMS.


We thank M. Stiles, and G. Bauer  for useful conversations, and J.C.
Slonczewski for useful suggestions including the notion that the
spin-torque can be calculated from the collinear spin currents. The
research at California State University Northridge was supported by
NSF grant DMR-00116566, US Army grant W911NF-04-1-0058, and NSF-KITP
grant PHY99-07949. Work at the University of Alabama was supported
by NSF MRSEC grant DMR 0213985, and by the INSIC-EHDR Program.


\end{document}